\documentclass[useAMS,usenatbib]{mn2e}
\usepackage[]{txfonts}
\def\simless{\mathbin{\lower 3pt\hbox
{$\rlap{\raise 5pt\hbox{$\char'074$}}\mathchar"7218$}}}   
\def\simmore{\mathbin{\lower 3pt\hbox
{$\rlap{\raise 5pt\hbox{$\char'076$}}\mathchar"7218$}}}   
\newcommand{\be}{\begin{equation}}
\newcommand{\ee}{\end{equation}}
\usepackage{graphicx}
\usepackage{epsfig} 
\usepackage{epstopdf}

\newcommand\apj{Astrophysical Journal}
\newcommand\apjl{Astrophysical Journal Letters}

\newcommand\aap{Astronomy \& Astrophysics}

\newcommand\nat{Nature}

\newcommand\mnras{Monthly Notices of the Royal Astronomical Society}

\title[Neutrino annihilation and GRB jets]
{Testing the neutrino annihilation model for launching GRB jets}

\author[Mingbin Leng and Dimitrios Giannios]
{Mingbin Leng$^{1}$\thanks{E-mail: mleng@purdue.edu  (ML)}
  and  Dimitrios Giannios$^1$\thanks{E-mail: dgiannio@purdue.edu  (DG)}\\
$^{1}$Department of Physics and Astronomy, Purdue University, 525 Northwestern
Avenue, West Lafayette, IN 47907, USA}

\begin{document}
\date{Received / Accepted}
\pagerange{\pageref{firstpage}--\pageref{lastpage}} \pubyear{2013}

\maketitle

\label{firstpage}

\begin{abstract}
The mechanism behind the launching of gamma-ray-burst (GRB) jets remains debated
resulting in large uncertainty over the jet composition. Both
magnetohydrodynamical and
neutrino annihilation models have been proposed for 
the energy extraction in a black hole/accretion-disc
central engine. In particular, for the extreme accretion rates $\dot M\sim 0.1-1$~
M$_\odot$s$^{-1}$ expected for bursts of duration $T\lesssim 100$~s, 
the disc can be an efficient neutrino emitter. Neutrino-antineutrino annihilation 
results in an energy deposition rate at the jet that can, 
in principle, account for the burst's energetics. Recent discoveries of X-ray flares hours after
the burst and of ultra-long GRBs suggest that GRB activity can last 
for $\sim 10^4$~s or longer. These long-lived events have fluence 
similar to that of classical GRBs. In view of these findings, we
re-evaluate the neutrino annihilation model. 
We derive the maximum possible energy of a neutrino-powered jet as 
a function of the burst duration and show that the available energy drops fast 
for longer bursts. For a standard choice of the parameters, 
the model falls short by three to four orders of magnitude in 
explaining the observed energetics of events that last longer than $\sim 10^3$~s.
\end{abstract} 
  

\section{Introduction} 
\label{intro}

 Gamma-ray bursts (GRBs) are powerful cosmic explosions of
characteristic duration of seconds. 
Their duration distribution is bimodal. Bursts with duration $T\lesssim
1$~s ($T\gtrsim 1$~s) are referred to as short (long) GRBs
(Kouveliotou et al. 1993).
Long GRBs are considered to come from the collapse of the
core of Wolf-Rayet stars (Woosley 1993) as demonstrated by their
association with Type Ic supernovae (Galama 1998; Stanek et al. 2003). Short GRBs are probably
result of merger of compact binaries (Eichler et al. 1989), 
though the observational evidence  for the nature of the 
progenitor remains sparse. 

Long GRBs have characteristic duration of $\sim 1-100$~s and
isotropic equivalent, gamma-ray release of $E^{\gamma}_{\rm iso}\sim
10^{52}-10^{54}$~erg (see, e.g., Bloom et al. 2003). However,
recently a new population of very long GRBs has been claimed
(including GRBs 101225A, 111209A, 121027A; Th\"one et al. 2011; Gendre
et al. 2013; Levan et al. 2014). These bursts last for $T\sim 10^4$~s
and, given their luminosity of 
$L^{\gamma}_{\rm iso}\sim 10^{49}-10^{50}$~erg $s^{-1}$, they have 
energy release similar to that of other powerful GRBs. Hereafter, we refer to these
bursts as ''ultra-long GRBs'' (Gendre et al. 2013). The host galaxies 
of ultra-long GRBs are
suggestive of massive star progenitors (Levan et al. 2014).\footnote
{Note that a stellar tidal disruption at a galactic centre
is another possibility for the origin of ultra-long GRBs.} 
The very long duration of these events 
can be accounted for by the extended size of the 
progenitor at core collapse (Woosley \& Heger 2012). 
Furthermore, GRBs exhibit powerful flares hours after the burst 
(Nousek et al. 2006; Zhang et al. 2006). Because of their fast
rise and decay time-scales, these flares are also 
believed to be powered by activity of the central engine
hours after the core collapse [see, however, Giannios (2006) 
for an alternative interpretation].

The energy source of the GRB is either the rotational energy of a
strongly magnetized proto-neutron star (millisecond magnetar; Usov 1992)
or the gravitational energy released during the accretion process to 
a newly-formed, a few solar mass black hole (Woosley 1993). 
Arguably, a black hole offers cleaner environment for the launch of a relativistic jet, though 
the protomagnetar model is viable alternative for long-duration
GRBs\footnote{Several seconds 
after its formation, the neutron star cools sufficiently 
to allow for relativistic winds to be launched from its surface 
(e.g., Metzger et al. 2011)}.

In the black hole scenarios, the jet may be launched by the
Blandford-Znajek mechanism, provided that sufficiently
strong magnetic fields thread the black hole horizon
(Blandford \& Znajek 1977).
Such a jet is expected to be magnetically dominated. 
It remains to be demonstrated that the magnetic
flux through the collapsing star is sufficient to power the jet
(Komissarov \& Barkov 2009). 

An alternative to magnetically-driven jets is that of
energy deposition through neutrino-antineutrino annihilation 
at the polar region of the black hole\footnote{Hereafter, we refer to
  neutrinos and antineutrinos collectively as neutrinos.}. Since the
newly formed black hole grows by several solar masses on a time-scale
of seconds,  the accretion rate  ($\dot{M}\sim 0.1- 1 M_\odot
/s$) is extreme. Under these conditions, the accretion disc
is a powerful source of neutrinos (Popham et al. 1999; 
Ruffert \& Janka 1999; Birkl et al. 2007; Chen \& Beloborodov 2007; Zalamea
\& Beloborodov 2011; hereafter ZB11). 
A fraction of these neutrinos annihilate into $e^{\pm}$ pairs, producing the
fireball that ultimately powers the GRB. 

The accretion rate  $\dot M$ on to the black hole is very uncertain since 
it depends on, among other things, the structure of the precollapse
star, its rotation profile, and the mass loss from the accretion disc
through outflows. 
Nevertheless, we show here that one can place very general limits on the
effectiveness of the neutrino annihilation mechanism. 
The efficiency with which neutrinos annihilate 
depends sensitively  on the accretion rate $\dot M$ and
the mass of the black hole $M_{\rm BH}$: $P_{\nu \bar \nu}\sim \dot M^{9/4}M_{\rm BH}^{-3/2}$
(ZB11). 
Therefore, by the time the mass of the central object roughly doubles,
the efficiency drops. Because of the finite mass available to be
accreted, It is clear that this mechanism cannot operate for
arbitrarily long time {\it while} maintaining a high accretion rate at
the black hole. Since the accretion rate is $\dot M\sim M_{\rm
  BH}/T$, the total energy available to power the jet is 
$E_{\nu \bar \nu}=P_{\nu \bar \nu}T\propto T^{-5/4}$, i.e., the bursts with the
longest duration cannot be as energetic as the shorter duration
ones. In this work, we demonstrate that the neutrino annihilation model potentially 
provides sufficient energy to power events of duration
$T\sim 1-100$~s but it falls short by $\sim 3-4$ orders of magnitude 
to account for the energetics of the longer observed bursts and the
X-ray flares that follow GRBs.
 
The structure of this Letter is the following. Section 2 summarizes the
main aspects of the neutrino annihilation model and its predictions 
for the burst luminosity as a function of its duration. Section 3 is
devoted to the comparison of the model with the various types of GRBs.
Implications from our findings are discussed in Section 4. 

\section{Neutrino-powered GRB flows}

For GRB relevant parameters the accretion disc feeds the black hole 
at a rate many orders of magnitude above the photon Eddington limit.
Such a disc cannot radiate away its excess energy with photons 
and, in the absence of an effective cooling mechanism, it is expected 
to be geometrically thick. For accretion rates approaching $\sim
1M_\odot$ s$^{-1}$, however, the density and temperature  in the disc 
become sufficiently high for neutrino emitting mechanisms to turn on 
close to the black hole. A fraction of these neutrinos annihilate 
into electron-positron pairs
($\nu\overline{\nu}\longrightarrow e^{+}e^{-}$), depositing energy in
the polar region of the black hole, and potentially launching a
relativistic jet. The power of the jet depends sensitively on the spin
and mass of the black hole as well as the accretion
rate.

Fast spinning black holes allow for the inner edge of the disc to approach closer to the 
black hole horizon. The inner disc becomes denser and hotter. As a result, the
neutrino emissivity of the disc is greatly enhanced. 
Given that the progenitor stars that can successfully power GRBs are
fast rotators, fast spinning black holes are naturally expected from 
such collapsars (Heger, Langer \& Woosley 2000). 
 In this work, we consider a black hole with dimensionless spin
parameter of $a=0.95$
(the high spin case studied in detail by ZB11; see below).

The jet power also depends on the accretion rate to the black
hole $\dot M$ and its mass $M_{\rm BH}$. Increasing $\dot M$ the disc 
is denser and neutrino luminosity increases. The neutrino annihilation is a two 
body process and the energy deposition rate increases steeply 
with $\dot M$. Also the mechanism is more effective for smaller black
holes. The smaller size of the central engine results in more compact and hot
discs (for fixed $\dot M$), i.e., more effective neutrino emitters.
   
The calculation of the energy deposition rate due to neutrino 
annihilation requires a detailed
model for the structure of a neutrino-cooled disc as well as
general-relativistic ray tracing of the neutrino orbits (see ZB11). 
The jet power can be approximated by the following expression
(for dimensionless black hole spin $a=0.95$): 
\be
P_{\nu \bar\nu}\approx1.3\times 10^{52} \Big(\frac{M_{\rm BH}}{3M_{\odot}}\Big)^{-3/2}
\times  
\left\{
\begin{array}{rl}
\left(\frac{\dot{M}}{M_{\odot} s^{-1}}\right)^{9/4} & \
\dot{M}_{\rm ign}<\dot{M}<\dot{M}_{\rm trap}\\
\left(\frac{\dot{M}_{\rm trap}}{M_{\odot} s^{-1}}\right)^{9/4}& \
  \dot{M}\ge \dot{M}_{\rm trap}
\end{array} \right.
\rm erg \,s^{-1},
\ee
where  
 $\dot{M}_{\rm ign}=0.021M_{\odot}$s$^{-1} (\frac{\alpha}{0.1})^{5/3}, and
\dot{M}_{\rm trap}=1.8M_{\odot}$s$^{-1} (\frac{\alpha}{0.1})^{1/3}$.
Here $\alpha$ stands for the standard viscosity parameter (Shakura \&
Sunyaev 1973). For $\dot M<\dot{M}_{\rm ign}$, the efficiency drops very
fast since the disc is not effectively cooling via neutrino emission and the jet
power $P_{\nu \bar \nu}$ is less that than predicted in equation.~(1). Since the
`ignition' accretion rate $\dot{M}_{\rm ign} $ depends sensitively on the uncertain 
viscosity parameter $\alpha$,
for the purpose of this work, we use equation.~(1) for any $\dot
M<\dot{M}_{\rm trap}$ with the understanding 
that we provide only an upper limit on the efficiency of the neutrino
annihilation process when  $\dot M<\dot{M}_{\rm ign}$.\footnote{The approach
  here is to evaluate whether the {\it maximum} possible jet energy
  predicted by the model is adequate to explain observations.}

\subsection{The predicted jet power}

So far, we have expressed the jet power as a function of the black hole
mass $M_{\rm BH}$ and the accretion rate $\dot M$, treating 
$M_{\rm BH}$ and $\dot M$ as independent variables. 
However, this is not the case in core
collapse supernova where the black hole may grow substantially  as
a result of accretion during the GRB. On the other hand, in a compact
object merger, the jet power is probably limited by the mass of the 
accretion disc. We discuss these two cases separately.  

\subsubsection{GRBs from collapsars}

The typical duration of long GRBs burst of $1-100$~s
is comparable to the free-fall time-scale of the progenitor 
Wolf-Rayet star. Since Wolf-Rayets undergo substantial mass loss
during their evolution, just prior to collapse such a progenitor has a
mass of $M\sim 10-15M_\odot$ (Heger, Langer  \& Woosley 2000). 
Since at least several $M_\odot$ are ejected during the supernova explosion, 
the final mass of the black hole remnant is $M_f\lesssim 10 M_\odot$.
For fast enough rotation of the progenitor, an accretion disc is expected to form 
around the black hole $\sim$seconds after core collapse facilitating the jet
launching. Still several seconds later 
the jet breaks through the collapsing star resulting in the GRB trigger.
Let $M_i \gtrsim 3 M_{\odot}$ be the mass of the black hole at jet
breakout/GRB trigger (MacFadyen \& Woosley 1999). 
For a Wolf-Rayet progenitor, the black
hole undergoes modest increase in mass during the burst
($M_f\sim$a few $M_i$).
We will first estimate the jet power as a function of burst duration 
assuming that the mass of the black hole evolves little during the burst. 
As a second step, we proceed with a more general calculation that
takes into account for the growth of the black hole mass.

For an average accretion rate $\dot M$ during a burst, the mass of the black hole
evolves as $M_{\rm BH}\sim M_{i}+\dot{M}t$. Assuming that the accretion episode
lasts time $T$ (e.g., the burst duration) during which the mass of the
black hole roughly doubles ($M_f=2M_i$), we have for the accretion rate: 
\be
\dot{M}\sim
\frac{M_f-M_i}{T}\sim\frac{M_i}{T}=0.3\frac{M_i}{3M_{\odot}}\Big(\frac{T}{10\rm{s}}\Big)^{-1}\rm
M_\odot s^{-1}.
\ee

This implies for the jet power that (see equation.~1)  
\be
P_{\nu \bar \nu}^I\sim 8.5\times 10^{50}\Big(\frac{M_i}{3M_{\odot}}\Big)^{3/4}\Big(\frac{T}{10\rm{s}}\Big)^{-9/4}
\rm erg\, s^{-1},
\ee
where the superscript $I$ stands for this, first, estimate of the jet power.

An alternative estimate of the jet power is to 
take into account that the mass of the black hole
evolves with time as implied by the accretion rate $\dot M=$d$M/$d$t$.
Equation.~(1) can then be rewritten as  
\be
\frac{{\rm d}M}{{\rm d}t}=\Big(\frac{P_{\nu \bar\nu}}{ 1.3 \times
  10^{52}\, \rm erg\, s^{-1}}\Big)^{4/9}\Big(\frac{M(t)}{ 3M_{\odot}}\Big)^{2/3}
 M_\odot \rm s^{-1}.
\ee
Assuming that the jet power is approximately constant during the
burst duration\footnote{This is an acceptable approximation as shown by
  the near linear increase with time of the cumulative count rates in GRB lightcurves
(McBreen et al. 2002)}, equation.~(4) can be integrated analytically. Setting
the integration time limits $t_1=0$ and $t_2=T$ and those of the
black hole mass $M_1=M_i$ and $M_2=M_f$, and 
solving for the jet power as function of the initial, final mass of the black hole and the burst
duration T results in:
\be
P_{\nu \bar\nu}^{II}=4.3\times
10^{51}M_i^{3/4}\Big((M_f/M_i)^{1/3}-1\Big)^{9/4}\Big(\frac{T}{10\rm{s}}\Big)^{-9/4}
\rm erg\, s^{-1}.
\ee 

For a final mass of the black hole that is twice as 
large as the initial mass, equation.~(5) gives
\be
P_{\nu \bar\nu}^{II} = 4.7 \times 10^{50}
\Big(\frac{M_{i}}{3M_\odot}\Big)^{3/4}
\Big(\frac{T}{10\rm{s}}\Big)^{-9/4} {\rm erg \,s^{-1}, \, for} \, M_f/M_i=2.
\ee
This expression is in reasonable agreement with equation.~(3). The factor of
$\sim 2$ difference in the predicted jet power comes for the fact that
equation.~(3) does not take into
account the drop in the efficiency because of the increase of the mass
of the black hole. In the following, unless otherwise specified, we 
keep expression (3) as a reference on the characteristic jet power predicted by
the neutrino annihilation model.

One can exploit equation.~(5) to derive a maximum possible jet power from the
collapse of a massive star (not necessarily a Wolf Rayet). 
By allowing a fairly large mass for the final black
hole of the remnant $M_f=40M_\odot$, the jet power becomes 
$P_{\nu \bar\nu}^{MAX}\sim 1.4\times
10^{52}(M_i/5M_\odot)^{3/4}(T/10s)^{-9/4}$ erg s$^{-1}$. 
This is an order of magnitude higher than ``standard'' 
estimate in equation.~(3) and may be more relevant for ultra-long GRBs.
If the star is fairly extended in size and remains very massive at the
moment of core collapse (e.g., as expected for blue supergiants with 
low mass-loss rate; Woosley \& Heger 2012), it can 
potentially power a burst of ultra-long duration. 
As we discuss below, even the maximum power predicted by the model
falls short by $\sim 2-3$ orders of magnitude in explaining the
observed properties of ultra-long GRBs.

\subsubsection{GRBs from compact object mergers}

The estimate (3) for the jet power is relevant for  GRBs associated 
with core collapse where the available matter
for accretion is similar to or exceeds that of the black hole. The merger of a
binary neutron star or of a black hole-neutron star system results in
a black hole surrounded by a light accretion disc $M_{disk}\lesssim
0.1 M_\odot$ (e.g., Ruffert \& Janka 1999). For the resulting accretion rate of 
$\dot{M}= 0.1(\frac{M_{disk}}{0.1M_{\odot}})(\frac{T}{1\rm{s}})^{-1}$
M$_\odot$ s$^{-1}$, the jet power is
\be
P_{\nu \bar\nu}^{\rm merger}=9.4\times 10^{49}\Big(\frac{M_{BH}}{2.5M_{\odot}}\Big)^{-3/2}\Big(\frac{M_{disc}}{0.1M_{\odot}}\Big)^{9/4}\Big(\frac{T}{1\rm{s}}\Big)^{-9/4}
\rm erg \,s^{-1}.
\ee

\section{Comparison with observations}

In Fig. 1, we schematically show the observed gamma-ray luminosity $L_{obs}^{\gamma}$ 
of various types of GRBs versus their observed duration $T$
(for a similar sketch see Levan et al. 2014). 
Long GRBs, short GRB as well as ultra-long GRBs and 
the X-ray flares that follow GRBs are shown. Long duration GRBs last for 
$T\sim 1-100$~s and have (isotropic equivalent ) luminosities up to 
$L_{obs}^{\gamma}\sim  10^{53}$~erg s$^{-1}$ and $E^{\gamma}_{\rm
  iso} \lesssim 10^{54}$~erg. Short GRBs typically
last a fraction of a second and reach luminosity similar to that of long
GRBs. Ultra-long GRBs (including GRBs 121027A, 101225A and 111209A)
last for $T\sim 10^3-10^4$~s and have luminosity in the  
$L_{obs}^{\gamma}\sim  10^{49}-10^{50}$~erg s$^{-1}$ range. 
X-ray flares take place $T_{\rm delay}\sim 100-10^5$~s after the 
GRB trigger and last for $T_f\sim 0.1 T_{\rm delay}\sim 10-10^4$~s (Chincarini et
al. 2007). Their fluence can approach that of GRBs but it is
typically $\sim$10 times smaller: $E_{\rm flare}^{\rm iso}\lesssim
3\times 10^{52}$~erg (Falcone et al. 2007). The typical peak luminosity of
the X-ray flares drops with time: $L_{\rm flare}^{\rm iso}\sim E_{\rm flare}^{\rm
  iso}/T_{\rm flare}$  (Chincarini et al. 2007).

To compare the jet power predicted by the model to the observed 
luminosity of GRBs, beaming and radiative efficiency corrections have
to be taken into account. 
The relationship between the true luminosity of the burst and the observed luminosity is
$ L_{\rm true}^{\gamma}=(\Omega/4\pi)L_{\rm obs}^{\gamma}$, where
$\Omega$ is the solid angle covered by the gamma-ray emission.
For jet opening angle $\theta$ and a symmetric, double jet system 
$\Omega\simeq 2\pi \theta^2$. Furthermore, if $\epsilon$ is
the radiative efficiency of the jet, one can compare the true jet power
$P_{\rm \nu\bar \nu}$ to the observed (isotropic equivalent) gamma-ray luminosity  
\be
L_{obs}^{\gamma}=\Big(\frac{2\epsilon}{\theta^{2}}\Big)P_{\nu \bar\nu}.
\ee

Using equation.~(3), we conclude that
\be
L_{obs}^{\gamma}=5.1\times 10^{52}\frac{\epsilon}{0.3}\Big(\frac{\theta}{0.1}\Big)^{-2}\Big(\frac{M_i}{3M_{\odot}}\Big)^{3/4}\Big(\frac{T}{10\rm{s}}\Big)^{-9/4}
\rm erg \,s^{-1},
\ee
where we adopt $\theta=0.1$ and $\epsilon=0.3$ as reference
values. The modelled-predicted jet luminosity is shown in Fig.~1. 

For completeness we also show in Fig.~1 the observed luminosity of gamma-ray
resulted from a binary merger event (see eq.~7):
\be
L_{obs}^{\gamma}\simeq 6\times
10^{51}\frac{\epsilon}{0.3}\Big(\frac{\theta}{0.1}\Big)^{-2}\Big(\frac{M_{\rm
  BH}}{2.5M_{\odot}}\Big)^{-3/2}\Big(\frac{M_{disc}}{0.1M_{\odot}}\Big)^{9/4}\Big(\frac{T}{1\rm{s}}\Big)^{-9/4}
\rm erg \,s^{-1}.
\ee

It is apparent that the energetics of the majority of 
long-duration bursts and short-duration GRBs can, 
in principle, be accounted by the model. 
Some tension exists between observations and theory 
for short GRBs with duration $\sim 1$~s as well long 
GRBs of $\sim 100$~s. In the case of short-duration GRBs, 
these conclusions rely on the presence of a rather massive disc 
$M_d\sim 0.1M_\odot$ around the merger product. 
If the disc mass is, instead, $M_d\sim 0.01M_\odot$ (e.g., Ruffert \&
Janka 1999), the disc is probably too light to power short-GRBs through 
neutrino annihilation. The mechanism is also less
effective for black hole/neutron star mergers. 
In that case, the final black hole has larger mass ($M_{\rm BH}\gtrsim
7 M_\odot$).      

\begin{figure}
\resizebox{\hsize}{!}{\includegraphics[]{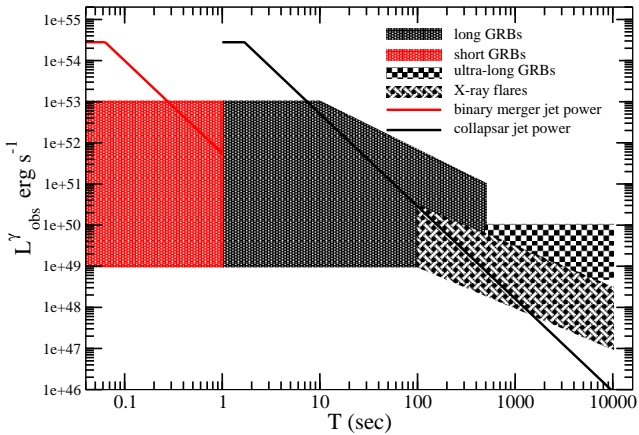}}
\caption[] {Schematic representation of the luminosity-duration
space occupied by various types of GRBs. The solid lines show the luminosity 
of a burst if powered by neutrino annihilation for jet opening angle
$\theta=0.1$ and radiative efficiency $\epsilon=0.3$ (red line for binary
mergers; see equation.~(10) and black line for core collapse GRBs; see equation.~(9)). While the model could
account for the majority of the bursts with duration $T\lesssim 100$~s,
it is challenged energetically to explain the longest events. 
\label{fig2}}
\end{figure}

The very long duration GRBs and X-ray flares
are hard to understand in the context of the neutrino annihilation
model. For events of duration $T\sim 10^4$~s,
our reference estimate for the jet observed luminosity is $L_{\rm
  obs}^{\gamma}\sim 10^{46}$~erg s$^{-1}$, i.e., short by $\sim$2 orders of magnitude 
to account for X-ray flares and $\sim 3-4$ orders of magnitude for ultra-long
GRBs, respectively.

\section{Discussion}

If ultra-long bursts are conclusively shown to be of core-collapse origin,
they pose a major challenge to neutrino annihilation models.
In such events, the accretion rate to the black hole has to be much lower 
than that during regular GRBs. Since the jet efficiency in this model 
depends steeply on the accretion rate ($P_{\nu \bar\nu}\propto \dot M^{9/4}$),
the model appears not to be energetically viable, for standard
choice of the parameters. Similar problems arise when applying the
model to the X-ray flares that follow a large fraction of GRBs.

One possibility is that some GRB jets are launched predominately 
by neutrino annihilation while others, the longer duration ones,  
by some other mechanism. However, there is no clear
observational evidence for different mechanisms involved in 
different bursts. Fig.~1 indicates that the GRB variety is likely
to originate from a continuum of core-collapse events powered 
by accretion to a few solar-mass
black hole where the duration is set by the size of the progenitor.
Furthermore, the observed spectral properties of GRBs and X-ray flares do
not show evidence for a sharp change from one 
type of source to the other (as might be expected for instance because
the jet composition is very different). Furthermore, relativistic jets
are universally observed from a broad range of black hole accretors
(e.g., blazars, microquasars) where neutrino annihilation is not 
of relevance. The simplest explanation is that
all GRB jets are driven, predominately, by a single mechanism,
unrelated to neutrino annihilation.

What if the reference values of the parameters we have adopted 
are not appropriate for the ultra-long GRBs. Can uncertainty in 
parameters have led us to underestimate the efficiency of the
mechanism by a factor of as large as $10^4$? The predicted jet power depends 
sensitively on the black hole spin and on the available mass to be
accreted. Furthermore, beaming corrections can be quite 
uncertain. For our reference model, we adopted a fast spinning black hole
of $a=0.95$. More extreme values of $a\sim 1$ may boost 
the efficiency of the mechanism by another factor of several. The mass
of the collapsing stars powering the ultra-long duration GRBs may be 
larger leading to black holes of $\sim 50M_\odot$. 
This raises the energy extracted by annihilation by another factor of $\sim 10$ (see
Section 2).   Finally, We have normalized the jet opening angle $\theta$ to 0.1 rad,
in accordance with typical expectation for GRBs. If ultra-long GRBs
have an opening angle of $\theta\lesssim 0.01$ rad, in combination with the
other possible boosting factors discussed above, neutrino annihilation 
might be able to account for observations. However,
such extreme beaming appears to us as unlikely. If true, such beaming will have profound 
implications for the true rates for these ultra-long GRBs. 

The neutrino annihilation model predicts a very specific trend among
GRBs: the longer the duration, the less energetic the burst. From
equation.~(3), we find that the burst energy
$E_{\nu \bar\nu}\sim P_{\nu \bar\nu}T\simeq 7.6\times 10^{48}(T/10^4s)^{-5/4}$~erg. 
GRBs that last for $T\sim 10, \, 100,\,1000,\,10^4$~s might have true (corrected for beaming)
energy of $E\sim 4\times 10^{52}, \, 2\times 10^{51}, \,
1.3\times 10^{50}, \,8\times 10^{48}$~erg, respectively.  
One can look for such a trend in the data since it is possible to
estimate the true energy of GRBs. The jet opening angle can be
constrained by the timing of jet breaks while late time radio
observations can be used to perform burst calorimetry (Frail et al. 2004). 
Even if such methods are approximate, the predicted trend of long-duration
bursts being weaker than short ones is strong enough to be tested observationally.

\section*{Acknowledgement}
DG acknowledges support from the Fermi 6 cycle grant number 61122.

\end{document}